\documentclass[lettersize,journal]{IEEEtran}
\IEEEoverridecommandlockouts
\usepackage{amsfonts}
\usepackage{mathrsfs}
\usepackage{bbding}
\usepackage{pifont} 
\usepackage{array}
\usepackage{amsthm}
\newtheorem{proposition}{Proposition}

\usepackage{tikz, pgfplots}
 \pgfplotsset{compat=1.18}

\usepackage{textcomp}
\usepackage{stfloats}
\usepackage[hidelinks]{hyperref}
\usepackage{booktabs}
\usepackage{url}
\usepackage{verbatim}
\usepackage{graphicx}
\usepackage{xcolor}
\usepackage{subfigure}
\usepackage{algorithm}
\usepackage{algpseudocode}
\usepackage{amsmath}
\usepackage{amssymb}
\usepackage{hyperref}
\usepackage{overpic, scalerel}
\usepackage{bm}
\usepackage{epstopdf}
\hypersetup{
    colorlinks=true,     
    linkcolor=blue,      
    citecolor=blue,      
    urlcolor=blue,       
    pdfborder={0 0 0}    
}
\usepackage{orcidlink}

\usepackage{soul}

\usepackage{cite}

\title{S-R$^3$: Secure and Robust RIS-Assisted RSMA under Finite-Resolution Phase Shifts}
\author{Ibrahima Sory Diallo, Elhadj Moustapha Diallo$^{_{\orcidlink{0009-0000-4860-6253}}}$, Wang Mi, Mamadou Aliou Diallo$^{_{\orcidlink{0009-0009-6860-8841}}}$, Amayika Kakati$^{_{\orcidlink{0009-0004-8315-1708}}}$

\thanks{I. S. Diallo is Department of Computer Science Engineering, Ahmadou Dieng University, Coyah 130, Guinea.}
\thanks{E. M. Diallo is with Ningbo Yinci Technology Co., Ltd. and Beijing SQUID Quantum Technology Co., Ltd., Beijing 100083, China.}
\thanks{W. Mi is with Beijing Milestone Technology Development Co.. Ltd., Beijing 100083, China.}
\thanks{M. A. Diallo and A. Kakati are with the School of Communications and Information Engineering, Chongqing University of Posts and Telecommunications, Chongqing, China}
\thanks{\raggedright Corresponding author: Elhadj Moustapha Diallo,}
\thanks{Email: L202110011@stu.cqupt.edu.cn.}
}

\begin{document}
\pagenumbering{arabic}

\maketitle
\bstctlcite{IEEEexample:BSTcontrol}

\begin{abstract}
This paper investigates a reconfigurable intelligent surface (RIS)-assisted rate-splitting multiple access (RSMA) downlink with imperfect eavesdropper channel state information and finite-resolution RIS phases. The common stream carries public information, while user-specific private streams carry confidential information. For a conservative security assessment, each eavesdropper is allowed to remove the public stream and is given the other private streams as side information. We formulate a robust design of the transmit precoders, isotropic artificial noise, common-rate allocations, and RIS phases that balances confidential and public throughput under rate and power constraints. Bounded errors in the direct and cascaded eavesdropper channels yield an effective-channel uncertainty bound valid across RIS phase updates. To address the resulting nonconvex problem, we construct conservative leakage bounds and convex inner transmit updates, initialize through a feasibility procedure, and search finite artificial-noise and RIS-phase candidates with explicit feasibility checks. For a fixed artificial-noise power and RIS configuration, the feasible inner iterates have a nondecreasing objective that converges in value, and their successive changes vanish. Accepted outer updates remain feasible and are finite in number.
Our code is available at \href{https://github.com/ABM-ADAM/RIS_RSMA-secure}{https://github.com/ABM-ADAM/RIS$\_$RSMA-secure}.
\end{abstract}
\begin{IEEEkeywords}
Artificial noise, finite-resolution phase shifts, physical-layer security,
rate-splitting multiple access (RSMA), reconfigurable intelligent surface
(RIS), robust optimization.
\end{IEEEkeywords}
\section{Introduction}
\label{sec:sec1}
Reconfigurable intelligent surfaces (RISs) shape reflected propagation paths in a multiuser downlink, affecting the channels of both legitimate users and potential eavesdroppers. Secure RIS-assisted transmission therefore calls for jointly designing the base-station precoders, artificial noise (AN), and RIS phases~\cite{yu2020robust_ris_security,peng2023robust_ris_security}. Practical RIS elements provide a finite set of phase states, whereas continuous-phase control is an idealization~\cite{wu2020discrete_ris}. Furthermore, the eavesdropper channels may be imperfectly known at the transmitter~\cite{yu2020robust_ris_security}. A security formulation should account for both finite phase resolution and eavesdropper-channel uncertainty.

Rate-splitting multiple access (RSMA) splits user messages into common and private parts: the common stream is decoded by all served users before each user decodes its own private stream. This structure can also accommodate multicast and unicast traffic in the same transmission~\cite{mao2019rs_unicast_multicast}. We assign different security requirements to the two streams: the common stream carries \emph{public} message parts, whereas each private stream carries the \emph{confidential} part of one user's message. Thus, the public common rate contributes to delivered throughput but never to confidential throughput. Unlike RIS--RSMA formulations that allocate a confidential common rate~\cite{wang2025_ris_rsma}, our security model allows an eavesdropper to know and cancel the public common stream. The RIS, transmit precoders, and AN are consequently designed to protect each confidential private stream under that assumption.

This paper studies that mixed-traffic downlink with bounded eavesdropper-channel errors and finite-resolution RIS phases. The objective balances a weighted sum of guaranteed confidential private rates against delivered public common rate, subject to explicit minimum-rate and transmit-power constraints. Because the resulting problem couples robust leakage bounds, multiuser decoding, AN, and discrete phase choices, both the security meaning of a rate and the scope of an algorithmic guarantee require careful specification.

\subsection{Related Works}

RIS-assisted secure transmission has been investigated in several aerial and multiuser settings. Joint transmit beamforming, RIS phases, and unmanned aerial vehicle (UAV) trajectory are optimized in~\cite{10494543}, while the impact of UAV jitter and imperfect hardware is addressed in~\cite{10075533}. More recently, hybrid RISs and multihop propagation have been studied for secure indoor--outdoor aerial millimeter-wave links~\cite{11265732}; secure content fetching in the presence of aerial and ground eavesdroppers is considered in~\cite{11162208}. These settings establish the relevance of propagation control and uncertain operating conditions, although their airborne geometry and traffic objectives differ from the fixed, narrowband downlink considered here.

Secure RSMA has also been studied in UAV networks through deep unfolding and reinforcement learning~\cite{10290852}, quality-of-experience-aware resource allocation~\cite{11432652}, and quantum deep reinforcement learning~\cite{11076787}. For RIS-assisted multiuser secrecy, the precoding study in~\cite{adam2023real} develops an alternating-optimization solution and an unfolded network for its original model. The closer RIS--RSMA work of Wang \emph{et al.}~\cite{wang2025_ris_rsma} optimizes a max--min secrecy objective with transceiver hardware impairments and an allocation of \emph{confidential} common rate. In contrast, the present common stream is public and may be known to the eavesdroppers. Its allocation cannot be inserted into a secrecy sum, and a comparison with~\cite{wang2025_ris_rsma} requires adapting the formulations to the same traffic and channel assumptions.

Learning-based methods address related but distinct RIS control problems: hierarchical reinforcement learning for terahertz massive MIMO~\cite{11180800}, multimodal reinforcement learning for an interference-limited RIS--UAV network~\cite{11076827}, and a discrete policy-gradient method for statistical quality of service in RIS-aided holographic MIMO~\cite{11131147}. A quantum-enhanced multi-RIS approach has also been studied for millimeter-wave Internet-of-Things networks using nonorthogonal multiple access~\cite{11494039}. These works illustrate different channel, traffic, or control models; none of their reported performance numbers is used as a surrogate for the robust public-common/private-confidential problem posed here.

\begin{figure*}
    \centering
    \includegraphics[width=7 in]{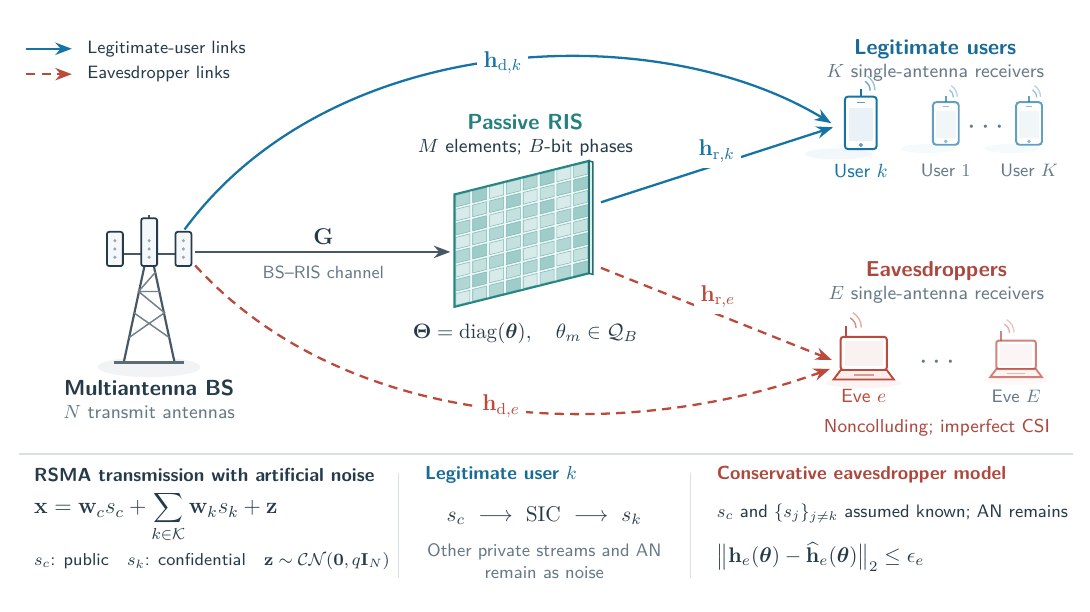}
    \caption{RIS-assisted RSMA downlink with noncolluding eavesdroppers. ChatGPT-6 Astra is used to plot the figure.}
    \label{fig:sys}
\end{figure*}

\subsection{Motivations and Contributions}
The above literature motivates a formulation in which the security claim matches the information actually required to be secret. Counting a public common allocation as secrecy, or relying on other users' private streams to remain undecodable at an eavesdropper, would overstate the confidentiality provided by this model. A second issue is that RIS phase updates change the effective eavesdropper channel, so uncertainty defined only around one fixed RIS configuration need not remain valid after an update. Finally, finite phase resolution and coupled rate constraints call for an implementable procedure with guarantees stated for the problem it actually solves. The contributions are as follows:

\begin{itemize}
\item \emph{A traffic-consistent robust model:} We formulate a downlink with a public common stream, confidential private streams, isotropic AN, and a finite-resolution passive RIS. Eavesdroppers are allowed to cancel the public stream and, conservatively, are given the other private streams as side information. We bound direct-link and cascaded-channel estimation errors and derive an enclosing effective-channel uncertainty ball valid for every admissible RIS configuration. The public and confidential rate constraints and their objective weights remain separate.

\item \emph{A verifiable joint design procedure:} We upper-bound worst-case private-stream leakage and construct convex inner transmit updates for fixed AN power and RIS phases using rate minorization, leakage majorization, and a proximal term. A Phase~I procedure searches for a feasible starting point; finite AN-power candidates and quantized RIS coordinate changes are then evaluated with direct checks of the physical constraints before acceptance. The method returns a feasible conservative design whenever its initialization and subsequent numerical solves succeed.

\item \emph{Explicit guarantees and evaluation protocol:} We provide pseudocode and arithmetic-complexity scaling for the finite-budget procedure. For a fixed AN-power/RIS pair and a feasible initial point, the inner objective is monotone and converges in value, with successive steps vanishing under the stated proximal update. Accepted outer moves remain feasible and improve the objective by a prescribed positive threshold, so their number is finite. The numerical methodology separates confidential and public rates and calls for an adapted RIS--RSMA literature baseline evaluated on paired channel draws.
\end{itemize}

Section~\ref{sec:system_model} specifies the signal model and robust problem. Section~\ref{sec:solution} develops the solution and its guarantees, and Section~\ref{sec:sec4} describes the simulation methodology and results.

\section{System Model and Problem Formulation}
\label{sec:system_model}
\subsection{Network and Channel Model}
\label{subsec:network_model}

We consider the downlink of a reconfigurable intelligent surface (RIS)-assisted rate-splitting multiple access (RSMA) network, as depicted in Fig.~\ref{fig:sys}. A base station (BS) equipped with $N$ antennas serves $K$ single-antenna legitimate users in the presence of $E$ noncolluding, single-antenna eavesdroppers. The RIS has $M$ passive reflecting elements. The sets of legitimate users and eavesdroppers are denoted by $\mathcal K=\{1,\ldots,K\}$ and $\mathcal E=\{1,\ldots,E\}$, respectively.

The BS-to-RIS channel is $\mathbf G\in\mathbb C^{M\times N}$. For legitimate user $k$, let $\mathbf h_{d,k}\in\mathbb C^{N}$ and $\mathbf h_{r,k}\in\mathbb C^{M}$ denote the direct BS-to-user and RIS-to-user channels, respectively. Similarly, $\mathbf h_{d,e}\in\mathbb C^{N}$ and $\mathbf h_{r,e}\in\mathbb C^{M}$ denote the channels associated with eavesdropper $e$. A blocked direct link is represented by setting its corresponding direct-channel vector to zero; it is not assigned a nonzero line-of-sight component.

The RIS employs $B$-bit phase control with alphabet
\begin{equation}
 \mathcal Q_B=\left\{e^{\mathrm j2\pi\ell/2^B}:\ell=0,\ldots,2^B-1\right\}.
 \label{eq:ris_phase_alphabet}
\end{equation}
The RIS reflection coefficients are collected in
\begin{equation}
 \begin{aligned}
 \boldsymbol\theta
 &=\big[e^{\mathrm j\phi_1},\ldots,e^{\mathrm j\phi_M}\big]^{T},\\
 \boldsymbol\Theta&=\operatorname{diag}(\boldsymbol\theta),
 \quad \theta_m\in\mathcal Q_B.
 \end{aligned}
 \label{eq:ris_coefficients}
\end{equation}
For a fixed RIS configuration, the effective BS-to-receiver channels are
\begin{align}
 \mathbf h_k^{H}(\boldsymbol\theta)
 &=
 \mathbf h_{d,k}^{H}
 +\mathbf h_{r,k}^{H}\boldsymbol\Theta\mathbf G,
 && k\in\mathcal K,
 \label{eq:effective_user_channel}\\
 \mathbf h_e^{H}(\boldsymbol\theta)
 &=
 \mathbf h_{d,e}^{H}
 +\mathbf h_{r,e}^{H}\boldsymbol\Theta\mathbf G,
 && e\in\mathcal E.
 \label{eq:effective_eve_channel}
\end{align}
Both effective channels in \eqref{eq:effective_user_channel}--\eqref{eq:effective_eve_channel} have dimension $1\times N$. The channel generation model, including path loss, fading distributions, spatial correlation, and any line-of-sight components, will be specified separately in the simulation methodology. These statistical assumptions are not needed to define the instantaneous signal and optimization models below.

We assume that the BS and RIS controller know the legitimate-user channels during the considered transmission interval. For eavesdropper $e$, write its effective-channel row vector as
\begin{equation}
 \begin{aligned}
 \mathbf h_e^H(\boldsymbol\theta)
 &=\mathbf h_{d,e}^H+\boldsymbol\theta^T\mathbf B_e,\\
 \mathbf B_e&=\operatorname{diag}(\mathbf h_{r,e}^{*})\mathbf G
 \in\mathbb C^{M\times N}.
 \end{aligned}
 \label{eq:eve_cascaded_channel}
\end{equation}
The controller has estimates $\widehat{\mathbf h}_{d,e}$ and $\widehat{\mathbf B}_e$, while the underlying channel errors obey
\begin{equation}
 \|\mathbf h_{d,e}-\widehat{\mathbf h}_{d,e}\|_2
 \leq\epsilon_{d,e},\qquad
 \|\mathbf B_e-\widehat{\mathbf B}_e\|_F
 \leq\epsilon_{B,e}.
 \label{eq:primitive_uncertainty}
\end{equation}
Define the estimated effective channel by $\widehat{\mathbf h}_e^H(\boldsymbol\theta) =\widehat{\mathbf h}_{d,e}^H+ \boldsymbol\theta^T\widehat{\mathbf B}_e$. For every admissible $\boldsymbol\theta$, the effective-channel error satisfies
\begin{equation}
 \|\mathbf h_e(\boldsymbol\theta)-
 \widehat{\mathbf h}_e(\boldsymbol\theta)\|_2
 \leq\epsilon_e\triangleq\epsilon_{d,e}+\sqrt M\epsilon_{B,e}.
 \label{eq:uniform_effective_bound}
\end{equation}
Indeed, $\|\boldsymbol\theta\|_2=\sqrt M$ and $\|\boldsymbol\theta^T(\mathbf B_e-\widehat{\mathbf B}_e)\|_2 \leq\sqrt M\epsilon_{B,e}$. This bound holds uniformly when the RIS phases change. For compactness, define the enclosing uncertainty set
\begin{equation}
 \mathcal U_e\triangleq
 \{\boldsymbol\Delta\in\mathbb C^N:
 \|\boldsymbol\Delta\|_2\leq\epsilon_e\},\quad
 \mathbf h_e(\boldsymbol\theta)=
 \widehat{\mathbf h}_e(\boldsymbol\theta)+\boldsymbol\Delta_e.
 \label{eq:eve_uncertainty}
\end{equation}
The ball $\mathcal U_e$ can contain effective-channel errors that cannot arise from the primitive errors in \eqref{eq:primitive_uncertainty}; protecting against the entire ball is conservative. These radii are deterministic bounds, not Gaussian standard deviations. Setting both radii to zero gives perfect estimated eavesdropper CSI.

\subsection{RSMA Transmission and Legitimate-User Rates}
\label{subsec:rsma_signal}

Each user message is split into a public part and a confidential part. The public parts are jointly encoded into one common stream $s_c$, which every legitimate user decodes. The confidential part of user $k$ is encoded into its private stream $s_k$. The streams are independent and satisfy $\mathbb E[|s_c|^2]=\mathbb E[|s_k|^2]=1$. The BS transmits
\begin{equation}
 \mathbf x
 =
 \mathbf w_c s_c
 +\sum_{k\in\mathcal K}\mathbf w_k s_k
 +\mathbf z,
 \label{eq:transmitted_signal}
\end{equation}
where $\mathbf w_c,\mathbf w_k\in\mathbb C^N$ are the common and private precoders. The independently generated artificial noise (AN) obeys $\mathbf z\sim\mathcal{CN}(\mathbf 0,q\mathbf I_N)$, where $q\geq0$ is an optimized scalar. Consequently, the total average transmit-power constraint is
\begin{equation}
 \|\mathbf w_c\|_2^2
 +\sum_{k\in\mathcal K}\|\mathbf w_k\|_2^2
 +Nq
 \leq P_{\mathrm B}.
 \label{eq:bs_power}
\end{equation}
The AN in \eqref{eq:transmitted_signal} is a controllable transmission variable; it does not represent transmitter hardware distortion.

The signal received by legitimate user $k$ is
\begin{equation}
 y_k
 =
 \mathbf h_k^H(\boldsymbol\theta)\mathbf w_c s_c
 +\sum_{j\in\mathcal K}
   \mathbf h_k^H(\boldsymbol\theta)\mathbf w_j s_j
 +\mathbf h_k^H(\boldsymbol\theta)\mathbf z+n_k,
 \label{eq:received_user}
\end{equation}
where $n_k\sim\mathcal{CN}(0,\sigma_k^2)$ and $\sigma_k^2>0$. For compactness, define
\begin{equation}
 D_k(\boldsymbol\theta,q)
 =q\|\mathbf h_k(\boldsymbol\theta)\|_2^2+\sigma_k^2.
 \label{eq:user_noise_an}
\end{equation}

Each legitimate user first decodes $s_c$ while treating all private streams and AN as noise. Its achievable common-stream rate is
\begin{equation}
 R_{c,k}
 =
 \log_2\!\left(
 1+
 \frac{
 |\mathbf h_k^H(\boldsymbol\theta)\mathbf w_c|^2
 }{
 \sum_{j\in\mathcal K}
 |\mathbf h_k^H(\boldsymbol\theta)\mathbf w_j|^2
 +D_k(\boldsymbol\theta,q)
 }
 \right).
 \label{eq:common_user_rate}
\end{equation}
After decoding and removing $s_c$, user $k$ decodes $s_k$, treating the other private streams and AN as noise. Its achievable private-stream rate is
\begin{equation}
 \begin{aligned}
 R_{p,k}&=\log_2(1+\gamma_{p,k}),\\
 \gamma_{p,k}
 &=\frac{|\mathbf h_k^H(\boldsymbol\theta)\mathbf w_k|^2}
 {\sum_{j\ne k}|\mathbf h_k^H(\boldsymbol\theta)
 \mathbf w_j|^2+D_k(\boldsymbol\theta,q)}.
 \end{aligned}
 \label{eq:private_user_rate}
\end{equation}
The rates in \eqref{eq:common_user_rate}--\eqref{eq:private_user_rate} assume successful common-stream cancellation. The effect of imperfect successive interference cancellation is outside the present signal model.

Let $C_k\geq0$ be the portion of the common-stream rate assigned to user $k$. Because all legitimate users must decode the complete common stream, the common-rate allocation must satisfy
\begin{equation}
 \sum_{k\in\mathcal K} C_k
 \leq R_{c,i},
 \qquad i\in\mathcal K.
 \label{eq:common_decodability}
\end{equation}
Writing the constraint for every $i$ avoids the nonsmooth $\min_i R_{c,i}$ notation and is equivalent to requiring that the aggregate common rate be decodable by the weakest legitimate user.

\subsection{Eavesdropper Model and Confidential Rates}
\label{subsec:secrecy_model}

The common stream carries public information. We therefore allow an eavesdropper to know and remove $s_c$ when attempting to decode a confidential private stream. We also give it the other users' private streams as side information. This latter assumption gives each eavesdropper a stronger decoding capability than simply treating the other private streams as noise; it avoids relying on multiuser interference as the sole source of confidentiality.

For an effective eavesdropper channel $\mathbf h_e(\boldsymbol\theta)$, the resulting information-leakage rate associated with the private stream of user $k$ is bounded by
\begin{equation}
 R_{e,k}\big(\mathbf h_e(\boldsymbol\theta)\big)
 =
 \log_2\!\left(
 1+
 \frac{
 |\mathbf h_e^H(\boldsymbol\theta)\mathbf w_k|^2
 }{
 q\|\mathbf h_e(\boldsymbol\theta)\|_2^2
 +\sigma_e^2
 }
 \right),
 \label{eq:eve_private_rate}
\end{equation}
where $\sigma_e^2>0$ is the receiver-noise variance of eavesdropper $e$. In \eqref{eq:eve_private_rate}, the other information streams do not appear in the denominator because they are assumed known to the eavesdropper for this conservative leakage calculation. The eavesdroppers are noncolluding: each acts on its own observation.

With independent Gaussian signaling and stochastic wiretap encoding of each confidential private message, the following private-stream rate difference is an achievable confidential-rate bound under the stated noncolluding, genie-aided eavesdropper model:
\begin{equation}
 R_{s,k}
 =\left[R_{p,k}
 -\max_{e\in\mathcal E}
 R_{e,k}\big(\mathbf h_e(\boldsymbol\theta)\big)\right]^+,
 \label{eq:instantaneous_secrecy}
\end{equation}
where $[a]^+\triangleq\max\{a,0\}$. The public allocation $C_k$ is not included in \eqref{eq:instantaneous_secrecy}. Thus, $C_k$ may contribute to a user's overall delivered rate, but it must not be reported as confidential throughput.

Under the bounded uncertainty model \eqref{eq:eve_uncertainty}, define the worst-case leakage rate
\begin{equation}
 \overline R_{e,k}
 =
 \max_{e\in\mathcal E}
 \;
 \sup_{\boldsymbol\Delta_e\in\mathcal U_e}
 R_{e,k}\!\left(
 \widehat{\mathbf h}_e(\boldsymbol\theta)
 +\boldsymbol\Delta_e
 \right).
 \label{eq:worst_case_leakage}
\end{equation}
The corresponding robust confidential rate is
\begin{equation}
 \underline R_{s,k}
 =
 \left[R_{p,k}-\overline R_{e,k}\right]^+.
 \label{eq:robust_secrecy_rate}
\end{equation}
The supremum in \eqref{eq:worst_case_leakage} is taken over a bounded channel-error set. It is not a chance constraint, and no secrecy-outage probability is claimed.

\subsection{Robust Joint Precoding and RIS Design}
\label{subsec:problem_formulation}

Let $\alpha_k>0$ denote the confidential-rate weight of user $k$, and let $\lambda\geq0$ denote the public-throughput weight. We introduce $t_k\geq0$ as its designed confidential rate. To ensure that the common stream serves an actual traffic requirement, let $r_{k}^{\mathrm{pub}}\geq0$ denote user $k$'s minimum public rate. An optional minimum confidential-rate requirement is denoted by $r_{k}^{\mathrm{sec}}\geq0$. These are distinct requirements: public bits cannot satisfy a confidential-rate requirement.

The joint public-and-confidential throughput design is parameterized by $\lambda$. We maximize a scalarized objective, while always reporting the public and confidential components separately. For $\lambda=0$ the formulation reduces to the confidential-throughput objective; for $\lambda>0$ the common stream can carry more than its minimum public traffic requirement. A sweep over $\lambda$ traces the attained public--confidential trade-off, subject to the limits of the nonconvex solver. Let $\mathcal V\triangleq(\mathbf w_c,\{\mathbf w_k\},q, \boldsymbol\theta,\{C_k,t_k\})$. The design is
\begin{subequations}
\label{prob:robust_rsma}
\begin{align}
 \max_{\mathcal V}
 \quad&
 \sum_{k\in\mathcal K}\alpha_k t_k
 +\lambda\sum_{k\in\mathcal K} C_k
 \label{prob:robust_rsma_obj}\\
 \text{subject to}\quad&
 t_k
 \leq
 R_{p,k}
 -
 R_{e,k}\!\left(
 \widehat{\mathbf h}_e(\boldsymbol\theta)
 +\boldsymbol\Delta_e
 \right),
 \nonumber\\[-0.3em]
 &\forall k\in\mathcal K,\ e\in\mathcal E,\
 \boldsymbol\Delta_e\in\mathcal U_e,
 \label{prob:robust_rsma_secrecy}\\
 &
 \sum_{k\in\mathcal K}C_k
 \leq R_{c,i},
 \quad \forall i\in\mathcal K,
 \label{prob:robust_rsma_common}\\
 &
 C_k\geq r_k^{\mathrm{pub}},
 \quad
 t_k\geq r_k^{\mathrm{sec}},
 \quad \forall k\in\mathcal K,
 \label{prob:robust_rsma_qos}\\
 &
 \|\mathbf w_c\|_2^2
 +\sum_{k\in\mathcal K}\|\mathbf w_k\|_2^2
 +Nq
 \leq P_{\mathrm B},
 \label{prob:robust_rsma_power}\\
 &
 q\geq0,
 \qquad
 \theta_m\in\mathcal Q_B,\quad m=1,\ldots,M.
 \label{prob:robust_rsma_ris}
\end{align}
\end{subequations}

Constraint \eqref{prob:robust_rsma_secrecy} limits $t_k$ by the legitimate private rate minus the leakage rate for every eavesdropper and every channel in its uncertainty set. Consequently, any feasible $t_k$ is no greater than $\underline R_{s,k}$ in \eqref{eq:robust_secrecy_rate}. The term $\lambda\sum_k C_k$ rewards only public throughput; it is never interpreted as a secrecy rate. Constraints \eqref{prob:robust_rsma_common} and \eqref{prob:robust_rsma_qos} ensure common-stream decodability and distinguish public from confidential traffic. The remaining constraints impose the BS power budget, nonnegative isotropic AN power, and finite-resolution passive RIS operation.

Problem \eqref{prob:robust_rsma} is nonconvex. In particular, the legitimate and eavesdropper rates are coupled through the precoders, AN covariance, and RIS phases; the semi-infinite secrecy constraints span the channel-uncertainty sets; and the discrete RIS phase decisions are combinatorial. A solution method must therefore specify how it handles each of these features. Neither alternating optimization nor semidefinite relaxation alone establishes global optimality. The next section develops a tractable iterative method and states the guarantees that its individual updates support.

\section{Conservative Robust Transmit and RIS Design}
\label{sec:solution}
\subsection{A Uniform Upper Bound on Leakage}
\label{subsec:leakage_bound}
For a chosen public-throughput weight $\lambda\geq0$, we seek a feasible, possibly suboptimal solution of \eqref{prob:robust_rsma}. For each RIS configuration, define
\begin{equation}
 a_e(\boldsymbol\theta)=
 \big[\|\widehat{\mathbf h}_e(\boldsymbol\theta)\|_2-\epsilon_e\big]_+^2,
 \quad [x]_+\triangleq\max\{x,0\}.
 \label{eq:eve_an_lower_factor}
\end{equation}
The uniform error bound \eqref{eq:uniform_effective_bound} implies
\begin{align}
 |\mathbf h_e^H(\boldsymbol\theta)\mathbf w_k|
 &\leq |\widehat{\mathbf h}_e^H(\boldsymbol\theta)\mathbf w_k|
       +\epsilon_e\|\mathbf w_k\|_2,
 \label{eq:leakage_numerator_bound}\\
 \|\mathbf h_e(\boldsymbol\theta)\|_2^2
 &\geq a_e(\boldsymbol\theta).
 \label{eq:leakage_denominator_bound}
\end{align}
Since $q\geq0$ and $\sigma_e^2>0$, a uniform upper bound on the leakage rate is
Define $b_{e,k}(\boldsymbol\theta,\mathbf w_k) \triangleq|\widehat{\mathbf h}_e^H(\boldsymbol\theta)\mathbf w_k| +\epsilon_e\|\mathbf w_k\|_2$. Then
\begin{equation}
 \overline R_{e,k}^{\rm ub}
 =\log_2\!\left(1+
 \frac{b_{e,k}^2(\boldsymbol\theta,\mathbf w_k)}
 {\sigma_e^2+q a_e(\boldsymbol\theta)}\right).
 \label{eq:conservative_eve_rate}
\end{equation}
For every admissible primitive-channel error, $R_{e,k}(\mathbf h_e(\boldsymbol\theta))\leq \overline R_{e,k}^{\rm ub}$. The numerator and denominator bounds need not be attained by the same channel error. Therefore,
\begin{equation}
 t_k+\overline R_{e,k}^{\rm ub}\leq R_{p,k},
 \qquad k\in\mathcal K,\ e\in\mathcal E,
 \label{eq:sufficient_secrecy}
\end{equation}
are sufficient, and generally conservative, constraints for \eqref{prob:robust_rsma_secrecy}. In what follows, the \emph{conservative physical problem} means \eqref{prob:robust_rsma} with its semi-infinite secrecy constraints replaced by \eqref{eq:sufficient_secrecy}.

\subsection{Convex Transmit Update at Fixed AN Power and RIS Phases}
\label{subsec:fixed_pair_update}
Fix $q=q_0\geq0$ and $\boldsymbol\theta=\boldsymbol\theta_0$. Write $\mathbf h_k=\mathbf h_k(\boldsymbol\theta_0)$, $\widehat{\mathbf h}_e=\widehat{\mathbf h}_e(\boldsymbol\theta_0)$, and $a_e=a_e(\boldsymbol\theta_0)$. The precoder-power budget is $P_0=P_{\rm B}-Nq_0$; we exclude pairs with $P_0<0$. For given precoders, define
\begin{align}
 T_{c,k}&=|\mathbf h_k^H\mathbf w_c|^2
 +\sum_{j\in\mathcal K}|\mathbf h_k^H\mathbf w_j|^2
 +q_0\|\mathbf h_k\|_2^2+\sigma_k^2,
 \label{eq:common_total_power}\\
 T_{p,k}&=\sum_{j\in\mathcal K}|\mathbf h_k^H\mathbf w_j|^2
 +q_0\|\mathbf h_k\|_2^2+\sigma_k^2.
 \label{eq:private_total_power}
\end{align}
For scalar complex equalizers $g_{c,k}$ and $g_{p,k}$, the common and private mean-square errors (MSEs) are
\begin{align}
 e_{c,k}&=|g_{c,k}|^2T_{c,k}
 -2\operatorname{Re}\{g_{c,k}^*\mathbf h_k^H\mathbf w_c\}+1,
 \label{eq:common_mse}\\
 e_{p,k}&=|g_{p,k}|^2T_{p,k}
 -2\operatorname{Re}\{g_{p,k}^*\mathbf h_k^H\mathbf w_k\}+1.
 \label{eq:private_mse}
\end{align}
For auxiliary MSE weights $u_{c,k},u_{p,k}>0$, set
\begin{align}
 \widetilde R_{c,k}&=
 \frac{\ln u_{c,k}-u_{c,k}e_{c,k}+1}{\ln2},
 \label{eq:common_rate_lower}\\
 \widetilde R_{p,k}&=
 \frac{\ln u_{p,k}-u_{p,k}e_{p,k}+1}{\ln2}.
 \label{eq:private_rate_lower}
\end{align}
For arbitrary auxiliaries, $\widetilde R_{c,k}\leq R_{c,k}$ and $\widetilde R_{p,k}\leq R_{p,k}$. Both are value tight at the current precoders under the MMSE updates
\begin{align}
 g_{c,k}&=\frac{\mathbf h_k^H\mathbf w_c}{T_{c,k}},
 &u_{c,k}&=\frac{1}{e_{c,k}},
 \label{eq:common_mmse_update}\\
 g_{p,k}&=\frac{\mathbf h_k^H\mathbf w_k}{T_{p,k}},
 &u_{p,k}&=\frac{1}{e_{p,k}},
 \label{eq:private_mmse_update}
\end{align}
where the MSEs on the right use the corresponding MMSE equalizers. Holding these auxiliaries fixed makes $\widetilde R_{c,k}$ and $\widetilde R_{p,k}$ concave quadratic functions of the precoders.

Introduce real $v_{e,k},\beta_{e,k}\geq0$ and impose
\begin{align}
 v_{e,k}&\geq
 |\widehat{\mathbf h}_e^H\mathbf w_k|
 +\epsilon_e\|\mathbf w_k\|_2,
 \label{eq:leakage_amplitude_slack}\\
 v_{e,k}^2&\leq d_e\beta_{e,k},
 \quad d_e\triangleq\sigma_e^2+q_0a_e>0.
 \label{eq:leakage_sinr_slack}
\end{align}
These convex constraints imply $\overline R_{e,k}^{\rm ub}\leq\log_2(1+\beta_{e,k})$. The second inequality has a rotated second-order-cone (SOC) representation because $d_e$ is constant. If $q$ were also variable here, that convexity argument would not apply.

At fixed-pair iteration $n$, set $v_{e,k}^{(n)}$ and $\beta_{e,k}^{(n)}$ to their smallest feasible values at the current precoders. Concavity of $\log_2(1+\beta)$ gives the affine global upper bound
\begin{equation}
 \begin{aligned}
 \log_2(1+\beta)&\leq
 \mathcal L_{e,k}^{(n)}(\beta),\\
 \mathcal L_{e,k}^{(n)}(\beta)
 &\triangleq\log_2(1+\beta_{e,k}^{(n)})\\
 &\quad+\frac{\beta-\beta_{e,k}^{(n)}}
 {(1+\beta_{e,k}^{(n)})\ln2}.
 \end{aligned}
 \label{eq:leakage_log_majorizer}
\end{equation}
It is value tight at $\beta_{e,k}^{(n)}$. Let $\mathbf X$ collect the real and imaginary parts of all precoders divided by a fixed positive amplitude scale and the real variables $\{C_k,t_k\}$ divided by a fixed positive rate scale. This invertible normalization makes the proximal norm dimensionless. Define
\begin{equation}
 F(\mathbf X)=\sum_k\alpha_k t_k+\lambda\sum_k C_k,
 \qquad \rho>0.
 \label{eq:inner_objective_definition}
\end{equation}
The fixed-pair inner update is the following convex program:
\begin{subequations}
\label{prob:fixed_pair_convex}
\begin{align}
 \max_{\mathbf X,\{v_{e,k},\beta_{e,k}\}}\quad
 &F(\mathbf X)-\frac{\rho}{2}
 \|\mathbf X-\mathbf X^{(n)}\|_2^2
 \label{prob:fixed_pair_obj}\\
 \mathrm{s.t.}\quad
 &t_k+\mathcal L_{e,k}^{(n)}(\beta_{e,k})
 \leq\widetilde R_{p,k}, &&\forall k,e,
 \label{prob:fixed_pair_secrecy}\\
 &\sum_{j\in\mathcal K}C_j
 \leq\widetilde R_{c,k}, &&\forall k,
 \label{prob:fixed_pair_common}\\
 &C_k\geq r_k^{\rm pub},\quad
 t_k\geq r_k^{\rm sec}, &&\forall k,
 \label{prob:fixed_pair_qos}\\
 &\|\mathbf w_c\|_2^2+
 \sum_{k\in\mathcal K}\|\mathbf w_k\|_2^2
 \leq P_0,
 \label{prob:fixed_pair_power}\\
 &\eqref{eq:leakage_amplitude_slack},\quad
 \eqref{eq:leakage_sinr_slack}, &&\forall k,e.
 \label{prob:fixed_pair_soc}
\end{align}
\end{subequations}
Here $v_{e,k},\beta_{e,k}\geq0$ are imposed explicitly; $C_k,t_k\geq0$ follow from nonnegative rate floors. The proximal term affects only the update; reported performance always uses $F$, not the penalized objective. The problem is a convex optimization problem (maximization of a concave objective over convex constraints), and its quadratic and norm constraints admit an SOC formulation.

\subsection{Feasible Initialization and Discrete Outer Search}
\label{subsec:outer_search}
A positive public or secrecy floor can make an all-zero precoder initialization infeasible. For a fixed pair, use the following auxiliary \emph{Phase I} problem. At iterate $n$, construct the MMSE rates and leakage majorizers exactly as above and solve
\begin{subequations}
\label{prob:phase_one}
\begin{align}
 \min_{\mathbf X,\{v_{e,k},\beta_{e,k}\},s}\quad&s
 \label{prob:phase_one_obj}\\
 \mathrm{s.t.}\quad
 &t_k+\mathcal L_{e,k}^{(n)}(\beta_{e,k})
 \leq\widetilde R_{p,k}+s, &&\forall k,e,
 \label{prob:phase_one_secrecy}\\
 &\sum_jC_j\leq\widetilde R_{c,k}+s, &&\forall k,
 \label{prob:phase_one_common}\\
 &\eqref{prob:fixed_pair_qos},\quad
 \eqref{prob:fixed_pair_power},\nonumber\\
 &\eqref{prob:fixed_pair_soc},\quad s\geq0.
 \label{prob:phase_one_other}
\end{align}
\end{subequations}
This is convex and always admits a sufficiently large $s$ if the initial precoders satisfy the power bound. After every solve, make the leakage slacks tight again before forming the next majorizer. If $s=0$, direct evaluation certifies feasibility of the conservative physical problem at that pair. If a finite initialization budget ends at $s>0$, record ``no feasible start found''; this does not prove infeasibility.

Select a finite AN design grid
\begin{equation}
 \mathcal Q_q\subseteq[0,P_{\rm B}/N],\qquad 0\in\mathcal Q_q.
 \label{eq:an_grid}
\end{equation}
Grid points restrict the \emph{search}, not the physical AN variable. Exhaustive search evaluates at most $|\mathcal Q_q|2^{BM}$ pairs. For a larger RIS, a budgeted coordinate search tests alternate phases of one element and alternate AN grid values. Whenever $q$ or $\boldsymbol\theta$ changes, a fresh feasible start must be certified; the incumbent precoders need not remain feasible at the candidate. An outer move is accepted only if its directly evaluated conservative physical constraints are satisfied and its unpenalized objective improves by more than $\delta_{\rm out}>0$. Feasibility checks include the public rate floors, the exact legitimate rates, power, and \eqref{eq:sufficient_secrecy} for every $(e,k)$.

\subsection{Pseudocode, Complexity, and Convergence}
\label{subsec:convergence}
Algorithms~\ref{alg:feasible_start}--\ref{alg:outer_search} specify the update and acceptance tests. In the idealized statements below, ``verify'' means exact evaluation; a numerical implementation must use tolerances and report its largest constraint residual.

\begin{algorithm}[t]
\caption{Feasible start for a fixed $(q_0,\boldsymbol\theta_0)$}
\label{alg:feasible_start}
\begin{algorithmic}[1]
\Require Pair with $P_0\geq0$; total Phase I solve budget $T_{\rm I}$; one or more power-feasible precoder seeds
\Ensure A verified conservative feasible $\mathbf X$, or \textsc{NotFound}
\For{each seed while total budget remains}
  \State Set $C_k=r_k^{\rm pub}$, $t_k=r_k^{\rm sec}$; set $n=0$
  \While{total Phase I budget remains}
    \State Set $v_{e,k}^{(n)}$ and $\beta_{e,k}^{(n)}$ tightly at the current precoders
    \State Update the MMSE equalizers and weights by \eqref{eq:common_mmse_update}--\eqref{eq:private_mmse_update}
    \State Solve Phase I \eqref{prob:phase_one}; retain its physical variables as $\mathbf X^{(n+1)}$
    \If{direct evaluation certifies \eqref{eq:sufficient_secrecy}, common decoding, QoS, and power}
      \State \Return $\mathbf X^{(n+1)}$
    \EndIf
    \State Increment $n$ and the total Phase I solve count
  \EndWhile
\EndFor
\State \Return \textsc{NotFound}
\end{algorithmic}
\end{algorithm}

\begin{algorithm}[t]
\caption{Conservative fixed-pair transmit optimization}
\label{alg:fixed_pair}
\begin{algorithmic}[1]
\Require Fixed feasible $(q_0,\boldsymbol\theta_0,\mathbf X^{(0)})$; $\rho>0$; maximum iterations $T_{\rm S}$; tolerance $\varepsilon_{\rm S}$
\Ensure Verified feasible $\mathbf X^{\rm out}$ and its unpenalized value $F^{\rm out}$
\For{$n=0,\ldots,T_{\rm S}-1$}
  \State Update the MMSE auxiliaries at $\mathbf X^{(n)}$ by \eqref{eq:common_mmse_update}--\eqref{eq:private_mmse_update}
  \State Set $v_{e,k}^{(n)}$ and $\beta_{e,k}^{(n)}$ tightly; form \eqref{eq:leakage_log_majorizer}
  \State Solve the SOC-representable problem \eqref{prob:fixed_pair_convex} and obtain $\mathbf X^{(n+1)}$
  \State Evaluate $F(\mathbf X^{(n+1)})$ and all conservative physical constraints
  \If{verification fails}
    \State \Return the last verified incumbent and record a numerical failure
  \EndIf
  \If{$F(\mathbf X^{(n+1)})-F(\mathbf X^{(n)})\leq\varepsilon_{\rm S}$}
    \State \Return $(\mathbf X^{(n+1)},F(\mathbf X^{(n+1)}))$
  \EndIf
\EndFor
\State \Return $(\mathbf X^{(T_{\rm S})},F(\mathbf X^{(T_{\rm S})}))$
\end{algorithmic}
\end{algorithm}

\begin{algorithm}[t]
\caption{Budgeted AN-grid and finite-phase coordinate search}
\label{alg:outer_search}
\begin{algorithmic}[1]
\Require $\mathcal Q_q,\mathcal Q_B$; candidate budget $H$; $\delta_{\rm out}>0$; Phase I and inner budgets
\Ensure Best verified incumbent found, or \textsc{NotFound}
\State Run Algorithm~\ref{alg:feasible_start} at chosen seed pairs until a verified pair is found
\If{no verified pair is found} \State \Return \textsc{NotFound} \EndIf
\State Run Algorithm~\ref{alg:fixed_pair} at that pair; store the verified incumbent and value $F_{\rm inc}$
\Repeat
  \State Set $\mathrm{improved}\gets\mathrm{false}$
  \State Form the one-element phase and alternate-$q$ neighbors of the current incumbent
  \State Set $F_{\rm best}\gets F_{\rm inc}$ and best candidate to empty
  \For{each neighbor, subject to remaining budget $H$}
    \State Run Algorithm~\ref{alg:feasible_start} at the candidate pair
    \If{the candidate has a verified feasible start}
      \State Run Algorithm~\ref{alg:fixed_pair}; directly verify its physical constraints
      \If{verification succeeds and $F_{\rm cand}>F_{\rm best}+\delta_{\rm out}$}
        \State Store this candidate as best; set $F_{\rm best}\gets F_{\rm cand}$
      \EndIf
    \EndIf
  \EndFor
  \If{a best candidate was stored}
    \State Replace the incumbent; set $F_{\rm inc}\gets F_{\rm best}$ and $\mathrm{improved}\gets\mathrm{true}$
  \EndIf
\Until{$\mathrm{improved}=\mathrm{false}$ or budget $H$ is exhausted}
\State \Return the verified incumbent with public and confidential rates reported separately
\end{algorithmic}
\end{algorithm}

\paragraph{Arithmetic complexity.}
Let $L=|\mathcal Q_q|$ and let $T_{\rm I}$ and $T_{\rm S}$ bound the Phase I and inner iterations per pair. With the cascaded channel matrices precomputed, forming effective channels for a pair takes $O(MN(K+E))$ arithmetic operations; updating the legitimate MMSE quantities and evaluating all $EK$ leakage amplitudes takes $O(N(K^2+EK))$ in a direct implementation. A standard SOC lift has $O(NK+K^2+EK)$ real scalar and cone coordinates; denote the actual Newton/KKT system dimension by $D$ and the aggregate self-concordant barrier parameter by $\nu$. For a standard interior-point solve with suitable initialization, a conservative dense factorization accounting bound is $O(\sqrt{\nu}\,D^3\log(1/\varepsilon_{\rm ip}))$ arithmetic operations per solve; this excludes canonicalization and exploits no sparsity. This is a solver model, not an iteration-independent timing prediction. If $H$ pairs are tested, an upper accounting bound is
\begin{equation}
 \begin{aligned}
 O\big(&H[MN(K+E)+(T_{\rm I}+T_{\rm S})\\
 &\times(NK^2+\sqrt{\nu}D^3
 \log(1/\varepsilon_{\rm ip}))]\big).
 \end{aligned}
 \label{eq:overall_complexity}
\end{equation}
Exhaustive enumeration has $H=L2^{BM}$. A complete coordinate sweep tests at most $M(2^B-1)+(L-1)$ alternatives, although accepted moves and finite budgets determine the number of sweeps. The exponential factor in exhaustive RIS search rules out a general real-time complexity claim for that variant.

\begin{proposition}[Feasibility and convergence of the inner update]
\label{prop:inner_convergence}
Assume $P_{\rm B}<\infty$, all receiver-noise variances are strictly positive, and a fixed pair has a feasible starting point for the conservative physical problem. Suppose \eqref{prob:fixed_pair_convex} is solved exactly at each iteration with $\rho>0$ and the specified tight MMSE and leakage updates. Then every physical iterate is feasible for \eqref{prob:robust_rsma}; $F(\mathbf X^{(n)})$ is nondecreasing and converges to a finite value; and $\|\mathbf X^{(n+1)}-\mathbf X^{(n)}\|_2\to0$.
\end{proposition}

\begin{proposition}[Verified outer acceptance]
\label{prop:outer_convergence}
Under the same finite-power and positive-noise assumptions, suppose each tested pair is processed for a finite number of iterations and accepted only after exact physical verification and an improvement greater than $\delta_{\rm out}>0$. Every retained incumbent is feasible for \eqref{prob:robust_rsma}, and there are finitely many accepted outer moves even if the search repeatedly revisits pairs. With a finite candidate budget, the procedure terminates. It does not assert optimality over untested or unsuccessfully initialized pairs.
\end{proposition}
Proofs are given in Appendix~\ref{app:algorithm_proofs}. The first proposition proves objective-value convergence and vanishing physical step lengths; it does not, by itself, prove convergence of the iterates to a unique point or KKT stationarity. The outer result does not establish optimality for continuously varying $q$ or a globally optimal RIS phase configuration.

\section{Simulation Results}
\label{sec:sec4}

We evaluate the robust RIS--RSMA design on one $360$-kHz NR physical resource block (PRB) at $3.5$~GHz. The direct BS--terminal links use the 3GPP urban-microcell (UMi) LOS/NLOS path-loss and LOS-probability expressions \cite{3gpp_38901_v195}. The RIS--terminal links use the same expressions with the RIS treated as a $10$-m node, while the clear BS--RIS hop uses ITU-R P.525-5 free-space loss \cite{itu_p525_5}. The latter RIS--terminal construction, independent small-scale fading across links, and independent shadowing are explicit modeling assumptions; the generator is not a spatially consistent 3GPP channel simulator. The $20$-MHz carrier power is allocated uniformly to the modeled $12$-subcarrier PRB at $30$-kHz subcarrier spacing \cite{3gpp_38211_v188,3gpp_38104_v193}. Thus, a $30$-dBm carrier budget corresponds to $P_{\rm B}=18$~mW ($12.55$~dBm) on the PRB. Table~\ref{tab:simulation_parameters} gives the nominal settings; the sweep ranges are stated with the corresponding figures.

\begin{table*}[!t]
\caption{Nominal simulation parameters for the blocked-corridor case.}
\label{tab:simulation_parameters}
\centering
\small
\begin{tabular}{@{}p{0.47\textwidth}p{0.47\textwidth}@{}}
\toprule
\textbf{Parameter} & \textbf{Value} \\
\midrule
Carrier frequency; carrier bandwidth; subcarrier spacing & $3.5$~GHz (n78); $20$~MHz; $30$~kHz \\
Modeled resource; receiver noise figure & One $360$-kHz PRB; $7$~dB \\
BS antennas; served users; noncolluding eavesdroppers & $N=8$; $K=4$; $E=2$ \\
RIS elements and phase resolution & $M=256$ ($16\times16$); $B=2$ bits \\
BS and RIS positions (m) & $(0,0,10)$; $(40,0,10)$ \\
User positions (m) & $[70,120]\times[-10,10]\times\{1.5\}$ \\
Eavesdropper positions (m) & $[60,130]\times\bigl([-35,-22]\cup[22,35]\bigr)\times\{1.5\}$ \\
Conditioned direct/RIS link states & Users: NLOS/LOS; eavesdroppers: LOS/NLOS \\
BS--RIS channel; RIS element efficiency & LOS free-space link; $0.8$ \\
LOS/NLOS small-scale Rician factors; shadowing standard deviations & $5/0$; $4/7.82$~dB; BS--RIS Rician factor $10$ \\
Carrier BS power; per-PRB budget $P_{\rm B}$ & $30$~dBm; $12.55$~dBm ($18$~mW) \\
Per-PRB receiver noise power & $\sigma_k^2=\sigma_e^2=-111.44$~dBm \\
Primitive eavesdropper CSI-error fraction $\rho$ & $0.01$; $\epsilon_{d,e}=\rho\|\widehat{\mathbf h}_{d,e}\|_2$, $\epsilon_{B,e}=\rho\|\widehat{\mathbf B}_{e}\|_F$ \\
AN fraction $Nq/P_{\rm B}$ & $\{0,0.05,0.10,0.20,0.30\}$ \\
Minimum public/confidential rate per user & $r_k^{\rm pub}=0.1$; $r_k^{\rm sec}=0$~bit/s/Hz \\
Objective weights & $\alpha_k=1$; $\lambda=0.2$ \\
Screened candidates; Phase-I/inner update budgets & $12$; $6/4$ \\
Independent seeds per operating point & $100$ for each main case; $50$ for power; $40$ for other sweeps \\
\bottomrule
\end{tabular}
\end{table*}

\paragraph*{Comparison and reporting protocol.}
The matched controls are \emph{no AN} ($q=0$), \emph{fixed RIS} (all RIS phases zero), and \emph{no RIS} (the cascaded channel suppressed); the remaining variables are optimized by the same finite-budget implementation. Each plotted comparison uses common channel seeds and, except for Fig.~\ref{fig:rsma_requirement}, conditions on the draws for which all displayed methods return directly verified solutions. We use $2000$ bootstrap resamples of these paired draws for the reported $95\%$ intervals; the sample sizes and intervals are retained in the accompanying numeric files. A failed feasible-start search is recorded as \texttt{not\_found}, not as a certificate of physical infeasibility. The AO--SCA/SDR method of Wang \emph{et al.} \cite{wang2025_ris_rsma} addresses a different max--min secrecy formulation with a confidential common stream and hardware impairments. An adaptation of that method is not implemented in these experiments, so the present figures support comparisons with the three matched controls only.

\subsection{Confidential Rate, CSI Uncertainty, and RIS Aperture}

\begin{figure}[!t]
\centering
\includegraphics[width=\columnwidth]{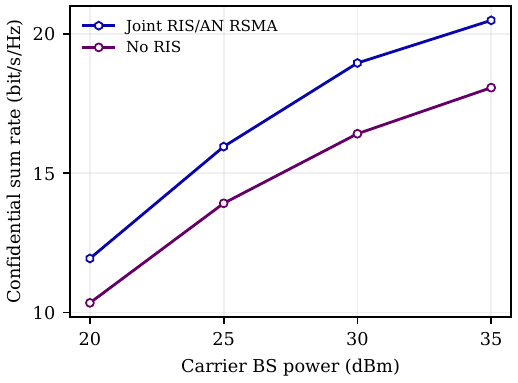}
\caption{Confidential sum rate versus carrier BS power in the blocked corridor. Power is allocated uniformly to one $360$-kHz PRB.}
\label{fig:rsma_power}
\end{figure}

Fig.~\ref{fig:rsma_power} shows that both methods gain from increased transmit power, while the joint design maintains a confidential-rate advantage over no RIS at every tested power. At $30$~dBm, the paired-draw means are $18.95$ and $16.41$~bit/s/Hz, respectively; their paired difference is $2.54$~bit/s/Hz with a $95\%$ bootstrap interval of $[2.03,3.14]$. The observed gain reflects the assumed reflective corridor and should not be extrapolated to arbitrary RIS placements.

\begin{figure}[!t]
\centering
\includegraphics[width=\columnwidth]{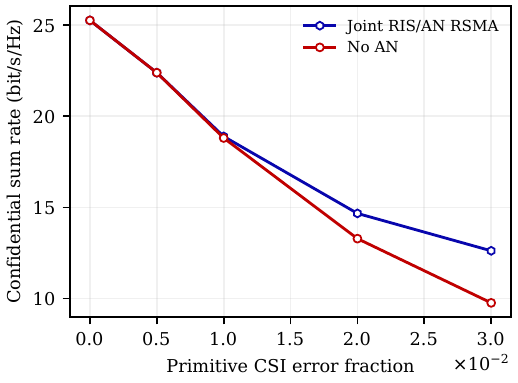}
\caption{Confidential sum rate versus primitive eavesdropper CSI-error fraction $\rho$. The no-AN control uses the same channel draws.}
\label{fig:rsma_uncertainty}
\end{figure}

The CSI-error fraction $\rho$ in Fig.~\ref{fig:rsma_uncertainty} affects both the direct and cascaded primitive eavesdropper channels. As $\rho$ increases from zero to $0.03$, the joint confidential rate decreases from $25.27$ to $12.62$~bit/s/Hz. AN provides essentially no gain at $\rho\leq0.005$ but increases the mean by $2.86$~bit/s/Hz over no AN at $\rho=0.03$ (paired interval $[1.94,3.81]$). Thus, the benefit of AN is conditional on uncertainty; it does not compensate for the overall loss due to a larger uncertainty set.

\begin{figure}[!t]
\centering
\includegraphics[width=\columnwidth]{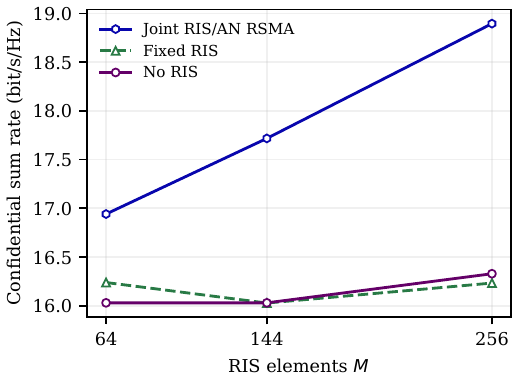}
\caption{Confidential sum rate versus RIS aperture size $M$ on nested physical channel draws. Fixed RIS uses zero phases.}
\label{fig:rsma_aperture}
\end{figure}

Fig.~\ref{fig:rsma_aperture} uses nested $8\times8$, $12\times12$, and $16\times16$ central subarrays of the same channel draw. The joint rate increases from $16.94$ to $18.89$~bit/s/Hz between $M=64$ and $256$, whereas the fixed-RIS rate remains near $16.2$~bit/s/Hz. The paired joint-minus-fixed-RIS gain consequently rises from $0.70$ to $2.66$~bit/s/Hz. In this geometry, a larger aperture is useful primarily when its phases are adapted, rather than merely increased in number.

\subsection{Public Traffic, Rate Distribution, and AN Selection}

\begin{figure}[!t]
\centering
\includegraphics[width=\columnwidth]{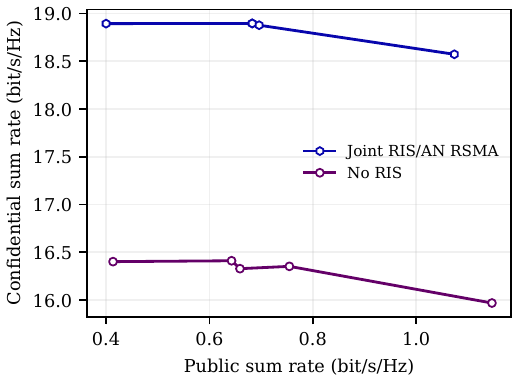}
\caption{Delivered public common rate and confidential private rate as the public-rate weight $\lambda$ varies.}
\label{fig:rsma_public_private}
\end{figure}

Fig.~\ref{fig:rsma_public_private} varies the public-rate weight over $\lambda\in\{0,0.1,0.2,0.5,1\}$. For the joint design, increasing the delivered public sum rate from $0.40$ to $1.07$~bit/s/Hz changes the confidential sum rate from $18.89$ to $18.57$~bit/s/Hz. The common stream contributes only to the horizontal coordinate: its public bits are excluded from confidential throughput. The limited movement of the points at intermediate weights is an outcome of the finite candidate search, rather than a claim that the plotted points trace a globally optimal rate region.

\begin{figure}[!t]
\centering
\includegraphics[width=\columnwidth]{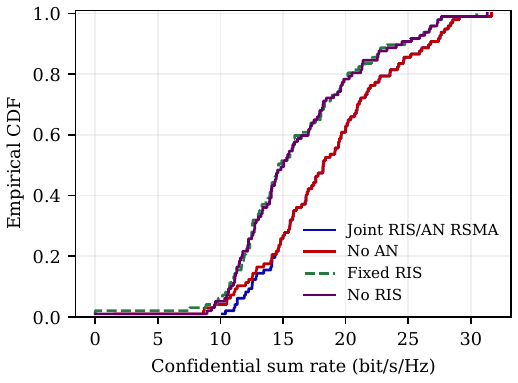}
\caption{Empirical CDF of confidential sum rate on the $97$ nominal draws verified by all four methods.}
\label{fig:rsma_ecdf}
\end{figure}

The empirical distributions in Fig.~\ref{fig:rsma_ecdf} use the intersection of $97$ verified nominal draws across all four methods. The paired mean gain of the joint design is $2.59$~bit/s/Hz over no RIS and $2.78$~bit/s/Hz over fixed RIS, with respective $95\%$ intervals $[2.10,3.12]$ and $[2.22,3.41]$. Its mean gain over no AN is only $0.25$~bit/s/Hz and occurs in a small subset of draws; the joint design exceeds no AN on $7$ of the $97$ paired draws. This distributional view avoids interpreting a small average AN gain as a typical per-channel gain.

\begin{figure}[!t]
\centering
\includegraphics[width=\columnwidth]{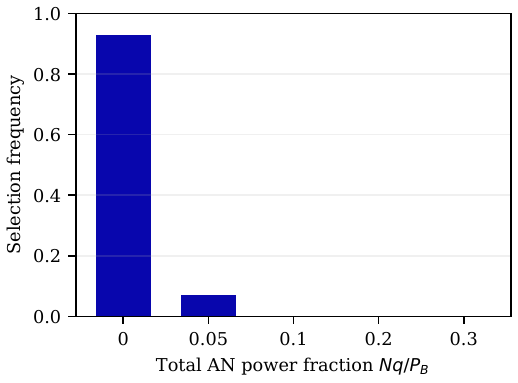}
\caption{Selection frequency of the total AN power fraction $Nq/P_{\rm B}$ on $99$ verified nominal draws.}
\label{fig:rsma_an_selection}
\end{figure}

Fig.~\ref{fig:rsma_an_selection} explains this behavior: of the $99$ verified nominal joint-design draws, $92$ select zero AN and $7$ select $Nq/P_{\rm B}=0.05$; none select a larger fraction. At $\rho=0.03$ in the uncertainty sweep, positive AN is instead selected on $23$ of $39$ verified draws. Hence, the AN grid is available to the optimizer, but its use is selective under the stated conservative leakage bound.

\subsection{Algorithm Behavior and Propagation Sensitivity}

\begin{figure}[!t]
\centering
\includegraphics[width=\columnwidth]{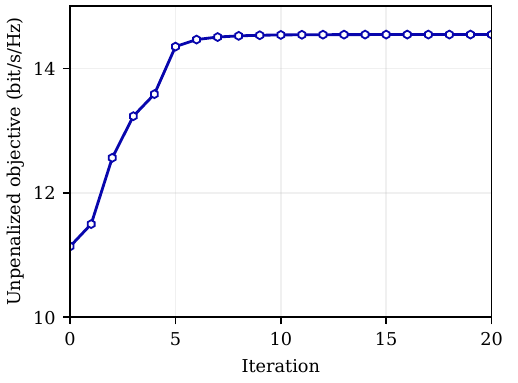}
\caption{Direct objective trace for a single fixed zero-phase RIS and $Nq/P_{\rm B}=0.05$ pair.}
\label{fig:rsma_inner_convergence}
\end{figure}

Fig.~\ref{fig:rsma_inner_convergence} reports one fixed-pair diagnostic: zero RIS phases, $Nq/P_{\rm B}=0.05$, and the prespecified median-rate nominal channel draw. The directly evaluated, unpenalized objective rises from $11.14$ to $14.54$~bit/s/Hz and levels off within approximately ten inner updates. This diagnostic uses a $20$-update budget, whereas the Monte Carlo campaign uses four inner updates; it neither depicts full alternating RIS/AN convergence nor establishes finite-run global optimality.

\begin{figure}[!t]
\centering
\includegraphics[width=\columnwidth]{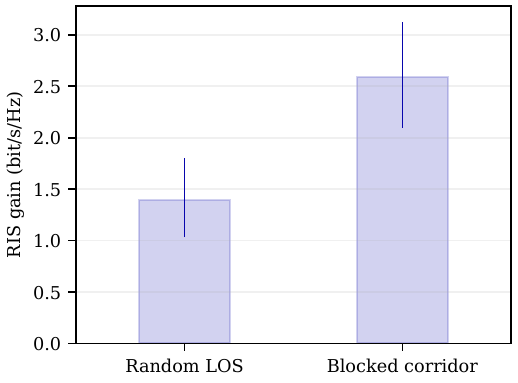}
\caption{Paired confidential-rate gain of joint RIS/AN over no RIS in the random-LOS and conditioned blocked-corridor cases; whiskers show $95\%$ bootstrap intervals.}
\label{fig:rsma_scenario}
\end{figure}

Fig.~\ref{fig:rsma_scenario} gives a paired joint-minus-no-RIS gain of $1.40$~bit/s/Hz under random UMi LOS states and $2.59$~bit/s/Hz in the conditioned blocked corridor. The respective bootstrap intervals are $[1.03,1.80]$ and $[2.10,3.12]$. The two cases also differ in terminal geometry and link-state conditioning; this comparison therefore measures the gain within each specified case, not the isolated causal effect of blockage.

\begin{figure}[!t]
\centering
\includegraphics[width=\columnwidth]{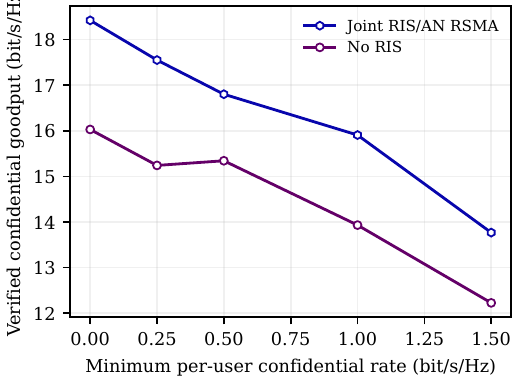}
\caption{Verified confidential goodput versus the minimum confidential rate per user; unsuccessful searches count as zero delivered service.}
\label{fig:rsma_requirement}
\end{figure}

Fig.~\ref{fig:rsma_requirement} imposes per-user minimum confidential rates from $0$ to $1.5$~bit/s/Hz. Here the ordinate is \emph{verified confidential goodput}: a search that does not return a verified design contributes zero service. At the largest requirement, the joint and no-RIS goodputs are $13.77$ and $12.23$~bit/s/Hz, while verified-solution counts fall to $30/40$ and $29/40$, respectively. The decline therefore reflects both the tighter service requirement and the reduced rate of successful feasible-start searches; it must not be described as a measured physical infeasibility probability.

\subsection{Search Time and Phase Resolution}

\begin{figure}[!t]
\centering
\includegraphics[width=\columnwidth]{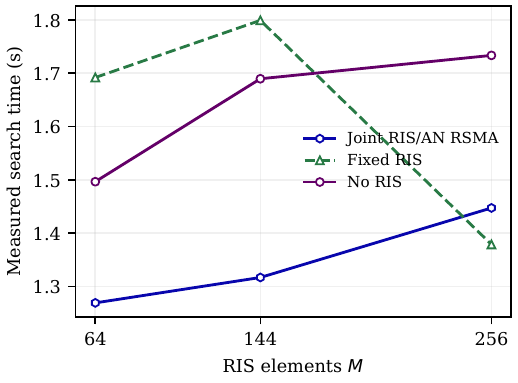}
\caption{Measured finite-budget search time versus RIS size for the joint, fixed-RIS, and no-RIS designs.}
\label{fig:rsma_runtime}
\end{figure}

Fig.~\ref{fig:rsma_runtime} reports wall-clock time for the finite-budget implementation. The joint-design mean rises from $1.27$ to $1.45$~s as $M$ increases from $64$ to $256$, but the fixed-RIS curve is nonmonotonic ($1.69$, $1.80$, and $1.38$~s). Variable solver paths and runtime variability preclude fitting an asymptotic complexity law to these three operating points; the values characterize this implementation and computing environment only.

\begin{figure}[!t]
\centering
\includegraphics[width=\columnwidth]{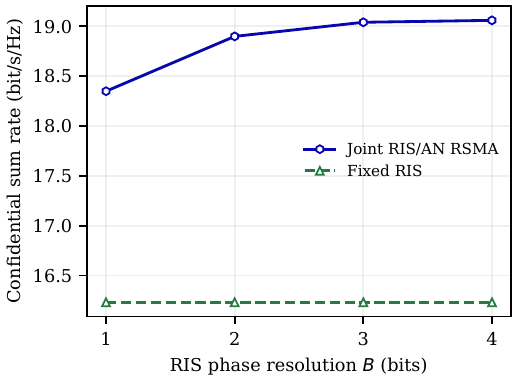}
\caption{Confidential sum rate versus RIS phase resolution $B$ on paired physical channel draws.}
\label{fig:rsma_bits}
\end{figure}

Finally, Fig.~\ref{fig:rsma_bits} shows that increasing the phase resolution from one to two bits raises the joint confidential rate from $18.35$ to $18.89$~bit/s/Hz; four bits yield $19.06$~bit/s/Hz. The fixed-zero-phase control is independent of $B$. Thus, two-bit phases capture most of the observed four-bit performance under this channel generator and candidate budget, without establishing a universal hardware-resolution optimum.

\section{Conclusion}
\label{sec:sec5}
This paper formulated a robust RIS-assisted RSMA downlink for mixed public and confidential traffic. The common stream carries public information, while confidential information is conveyed through private streams. The secrecy model allows each eavesdropper to cancel the common stream and treats the other private streams as known side information. Bounded errors in the direct and cascaded eavesdropper channels lead to an effective-channel uncertainty bound that remains valid as the RIS phases change.

We developed a conservative joint design of the transmit precoders, artificial noise, common-rate allocations, and finite-resolution RIS phases. The procedure combines feasible initialization, convex inner transmit updates, and a discrete outer search with direct feasibility checks. For a fixed artificial-noise power and RIS configuration, feasible inner iterates have a nondecreasing objective that converges in value; accepted outer updates remain feasible and are finite in number. These guarantees establish what the procedure delivers under the stated uncertainty model without implying global optimality.

\appendices
\section{Proofs of the Algorithmic Guarantees}
\label{app:algorithm_proofs}

\begin{proof}[Proof of Proposition~\ref{prop:inner_convergence}]
At iteration $n$, choose the MMSE auxiliaries at the physical incumbent. The resulting rate lower bounds equal the exact legitimate rates at $\mathbf X^{(n)}$. Choose the leakage slacks at their smallest feasible values: then $\beta_{e,k}^{(n)}=(|\widehat{\mathbf h}_e^H\mathbf w_k^{(n)}|+\epsilon_e\|\mathbf w_k^{(n)}\|_2)^2/d_e$. The affine leakage bound equals the conservative leakage rate at these slacks. Consequently, $(\mathbf X^{(n)},\mathbf v^{(n)},\boldsymbol\beta^{(n)})$ is feasible for \eqref{prob:fixed_pair_convex} whenever the physical incumbent is feasible for \eqref{eq:sufficient_secrecy}, common decoding, QoS, and power.

For any feasible point of \eqref{prob:fixed_pair_convex}, $\overline R_{e,k}^{\rm ub}\leq \log_2(1+\beta_{e,k})\leq \mathcal L_{e,k}^{(n)}(\beta_{e,k})$, whereas $\widetilde R_{p,k}\leq R_{p,k}$ and $\widetilde R_{c,k}\leq R_{c,k}$. Hence, its physical variables satisfy \eqref{eq:sufficient_secrecy} and the exact common, QoS, and power constraints. By induction, all iterates are feasible for the conservative physical problem and therefore also for \eqref{prob:robust_rsma}.

Use the old incumbent and tight slacks as a feasible point in the optimization at iteration $n$. Optimality of the new point gives
\begin{equation}
 F(\mathbf X^{(n+1)})-F(\mathbf X^{(n)})
 \geq \frac{\rho}{2}
 \|\mathbf X^{(n+1)}-\mathbf X^{(n)}\|_2^2\geq0.
 \label{eq:proximal_ascent}
\end{equation}
Because RIS phases belong to a finite alphabet, each $H_k\triangleq\max_{\boldsymbol\theta\in\mathcal Q_B^M} \|\mathbf h_k(\boldsymbol\theta)\|_2$ is finite. The power budget and positive noise give the uniform bound $R_{c,k},R_{p,k}\leq \log_2(1+P_{\rm B}H_k^2/\sigma_k^2)$. The secrecy inequalities imply $0\leq t_k\leq R_{p,k}$; common decoding gives $\sum_k C_k\leq\max_i R_{c,i}$. Thus $F$ has a finite upper bound. A nondecreasing bounded sequence of real values converges. Summing \eqref{eq:proximal_ascent} over $n$ bounds $\sum_{n\geq0}\|\mathbf X^{(n+1)}- \mathbf X^{(n)}\|_2^2$ by a finite constant, proving that its individual terms tend to zero.
\end{proof}

\begin{proof}[Proof of Proposition~\ref{prop:outer_convergence}]
Algorithm~\ref{alg:outer_search} stores only physically verified designs. Proposition~\ref{prop:inner_convergence} and the direct candidate check imply that every accepted incumbent satisfies the conservative constraints and hence the original robust constraints. Let $F_{\max}<\infty$ be the uniform objective bound from the preceding proof and $F_0$ be the initial incumbent value. After $J$ accepted moves, $F_J>F_0+J\delta_{\rm out}$; therefore $J<(F_{\max}-F_0)/\delta_{\rm out}$. Each sweep checks a finite set of alternatives and each candidate uses finite Phase I and inner budgets. Thus the number of accepted moves is finite, and a finite overall candidate budget also forces termination. A failed Phase I run conveys no infeasibility certificate and cannot be used to infer optimality.
\end{proof}

\bibliographystyle{IEEEtran}
\bibliography{rsma-ris}

@ARTICLE{10494543,
  author={Adam, Abuzar B. M. and Ouamri, Mohamed Amine and Wan, Xiaoyu and Muthanna, Mohammed Saleh Ali and Alkanhel, Reem and Muthanna, Ammar and Li, Xingwang},
  journal={IEEE Systems Journal}, 
  title={Secure Communication in UAV–RIS-Empowered Multiuser Networks: Joint Beamforming, Phase Shift, and UAV Trajectory Optimization}, 
  year={2024},
  volume={18},
  number={2},
  pages={1009-1019},
  doi={10.1109/JSYST.2024.3379456}}

@ARTICLE{10075533,
  author={Adam, Abuzar B. M. and Wan, Xiaoyu and Elhassan, Mohammed A. M. and Muthanna, Mohammed Saleh Ali and Muthanna, Ammar and Kumar, Neeraj and Guizani, Mohsen},
  journal={IEEE Transactions on Vehicular Technology}, 
  title={Intelligent and Robust UAV-Aided Multiuser RIS Communication Technique With Jittering UAV and Imperfect Hardware Constraints}, 
  year={2023},
  volume={72},
  number={8},
  pages={10737-10753},
  doi={10.1109/TVT.2023.3255309}}

@INPROCEEDINGS{10290852,
  author={Adam, Abuzar B. M. and Elhassan, Mohammed A. M.},
  booktitle={2023 International Conference on Communications, Computing, Cybersecurity, and Informatics (CCCI)}, 
  title={Enhancing Secrecy in UAV RSMA Networks: Deep Unfolding Meets Deep Reinforcement Learning}, 
  year={2023},
  volume={},
  number={},
  pages={1-5},
  doi={10.1109/CCCI58712.2023.10290852}}

@misc{adam2023real,
      title={Real-Time and Security-Aware Precoding in RIS-Empowered Multi-User Wireless Networks}, 
      author={Abuzar B. M. Adam and Mohamed Amine Ouamri and Mohammed Saleh Ali Muthanna and Xingwang Li and Mohammed A. M. Elhassan and Ammar Muthanna},
      year={2023},
      eprint={2401.00192},
      archivePrefix={arXiv},
      primaryClass={cs.IT},
      url={https://arxiv.org/abs/2401.00192}, 
}

@ARTICLE{11180800,
  author={Adam, Abuzar B. M. and Abdullah, Zaid and Chatzinotas, Symeon},
  journal={IEEE Transactions on Wireless Communications}, 
  title={Generative AI-Based Hierarchical DRL Framework for RIS-Assisted THz Massive MIMO Systems}, 
  year={2026},
  volume={25},
  number={},
  pages={4480-4495},
  doi={10.1109/TWC.2025.3611629}}

@ARTICLE{11265732,
  author={Kakati, Amayika and Li, Guoquan and Diallo, Elhadj Moustapha and Ouamri, Mohamed Amine and Khalid, Muhammad Imran and Adam, Abuzar B. M. and Li, Xingwang},
  journal={IEEE Transactions on Vehicular Technology}, 
  title={Hybrid RIS-Assisted Multi-Hop Indoor-Outdoor Secure Communication in AAV mmWave Networks}, 
  year={2026},
  volume={75},
  number={6},
  pages={10114-10129},
  doi={10.1109/TVT.2025.3636299}}

@INPROCEEDINGS{11076827,
  author={Diallo, Mamadou Aliou and Ouamri, Mohamed Amine and Khalid, Muhammad Imran and Komal, Asma and Diallo, Elhadj Moustapha and Adam, Abuzar. B. M.},
  booktitle={2025 12th IFIP International Conference on New Technologies, Mobility and Security (NTMS)}, 
  title={Multi-Modal Deep Reinforcement Learning Framework for Interference-Limited RIS-Assisted UAV Wireless Network}, 
  year={2025},
  volume={},
  number={},
  pages={80-85},
  doi={10.1109/NTMS65597.2025.11076827}}

@ARTICLE{11131147,
  author={Ouamri, Mohamed Amine and Adam, Abuzar B. M. and Elhassan, Mohammed A. M. and Moustapha Diallo, Elhadj and Li, Xingwang},
  journal={IEEE Communications Letters}, 
  title={Discrete Softmax Policy Gradient for Statistical QoS Provisioning in RIS-Aided Holographic MIMO Networks}, 
  year={2025},
  volume={29},
  number={11},
  pages={2516-2520},
  doi={10.1109/LCOMM.2025.3600872}}

@INPROCEEDINGS{11162208,
  author={Diallo, Elhadj Moustapha and Adam, Abuzar B. M. and Kakati, Amayika and Diallo, Mamadou Aliou and Amine Ouamri, Mohamed},
  booktitle={2025 IEEE International Conference on Communications Workshops (ICC Workshops)}, 
  title={Deep Learning-Based Secure Content Fetching in UAV-Aided and RIS-Assisted Under Coordinated Aerial and Ground Eavesdropping}, 
  year={2025},
  volume={},
  number={},
  pages={1544-1549},
  doi={10.1109/ICCWorkshops67674.2025.11162208}}

@INPROCEEDINGS{11432652,
  author={Ouamri, Mohamed Amine and Adam, Abuzar B. M. and Benallouche, Yacine and Gueroui, Mourad},
  booktitle={GLOBECOM 2025 - 2025 IEEE Global Communications Conference}, 
  title={Secure QoE-Aware UAV-Aided Rate-Splitting Multiple Access-Based Communications}, 
  year={2025},
  volume={},
  number={},
  pages={4565-4570},
  doi={10.1109/GLOBECOM59602.2025.11432652}}

@INPROCEEDINGS{11076787,
  author={Ouamri, Mohamed Amine and Adam, Abuzar B. M. and Elhassan, Mohammed A. M.},
  booktitle={2025 12th IFIP International Conference on New Technologies, Mobility and Security (NTMS)}, 
  title={Quantum Deep Reinforcement Learning for Secure Multi-User UAV RSMA Networks}, 
  year={2025},
  volume={},
  number={},
  pages={86-92},
  doi={10.1109/NTMS65597.2025.11076787}}

@ARTICLE{11494039,
  author={Diallo, Elhadj Moustapha and Adam, Abuzar B. M. and Elhassan, Mohammed A. M. and Hammoudeh, Mohammad and Muthanna, Mohammed Saleh Ali and Muthanna, Ammar},
  journal={IEEE Transactions on Consumer Electronics}, 
  title={Quantum-Enhanced Multi-RIS-Empowered Consumer Electronics mmWave IoT NOMA Networks}, 
  year={2026},
  volume={72},
  number={3},
  pages={5965-5973},
  doi={10.1109/TCE.2026.3687145}}

@techreport{3gpp_38901_v195,
  author      = {{3GPP}},
  title       = {{TR 38.901}: Study on channel model for frequencies from 0.5 to 100 {GHz}},
  institution = {3rd Generation Partnership Project},
  number      = {V19.5.0},
  month       = sep,
  year        = {2026},
  url         = {https://www.etsi.org/deliver/etsi_tr/138900_138999/138901/19.05.00_60/tr_138901v190500p.pdf}
}

@techreport{3gpp_38211_v188,
  author      = {{3GPP}},
  title       = {{TS 38.211}: {NR}; Physical channels and modulation},
  institution = {3rd Generation Partnership Project},
  number      = {V18.8.0},
  month       = feb,
  year        = {2026},
  url         = {https://www.etsi.org/deliver/etsi_ts/138200_138299/138211/18.08.00_60/ts_138211v180800p.pdf}
}

@techreport{3gpp_38104_v193,
  author      = {{{3GPP}}},
  title       = {{TS 38.104}: {NR}; Base Station ({BS}) radio transmission and reception},
  institution = {3rd Generation Partnership Project},
  number      = {V19.3.0},
  month       = feb,
  year        = {2026},
  url         = {https://www.etsi.org/deliver/etsi_ts/138100_138199/138104/19.03.00_60/ts_138104v190300p.pdf}
}

@techreport{itu_p525_5,
  author      = {{ITU-R}},
  title       = {Calculation of free-space attenuation},
  institution = {International Telecommunication Union},
  number      = {Recommendation ITU-R P.525-5},
  month       = nov,
  year        = {2024},
  url         = {https://www.itu.int/rec/R-REC-P.525}
}

@article{wang2025_ris_rsma,
  author  = {Kehao Wang and Dajin Li and Jialiang Zhang and Changzhen Li},
  title   = {Robust and secure transmission for {RIS}-assisted {RSMA} with transceiver hardware impairments},
  journal = {Physical Communication},
  volume  = {71},
  pages   = {102629},
  year    = {2025},
  doi     = {10.1016/j.phycom.2025.102629},
  url     = {https://doi.org/10.1016/j.phycom.2025.102629}
}

@article{yu2020robust_ris_security,
  author  = {Xianghao Yu and Dongfang Xu and Ying Sun and Derrick Wing Kwan Ng and Robert Schober},
  title   = {Robust and Secure Wireless Communications via Intelligent Reflecting Surfaces},
  journal = {IEEE Journal on Selected Areas in Communications},
  volume  = {38},
  number  = {11},
  pages   = {2637--2652},
  year    = {2020},
  doi     = {10.1109/JSAC.2020.3007043}
}

@article{peng2023robust_ris_security,
  author  = {Zhangjie Peng and Ruisong Weng and Cunhua Pan and Gui Zhou and Marco Di Renzo and A. Lee Swindlehurst},
  title   = {Robust Transmission Design for {RIS}-Assisted Secure Multiuser Communication Systems in the Presence of Hardware Impairments},
  journal = {IEEE Transactions on Wireless Communications},
  volume  = {22},
  number  = {11},
  pages   = {7506--7521},
  year    = {2023},
  doi     = {10.1109/TWC.2023.3252046}
}

@article{wu2020discrete_ris,
  author  = {Qingqing Wu and Rui Zhang},
  title   = {Beamforming Optimization for Wireless Network Aided by Intelligent Reflecting Surface With Discrete Phase Shifts},
  journal = {IEEE Transactions on Communications},
  volume  = {68},
  number  = {3},
  pages   = {1838--1851},
  year    = {2020},
  doi     = {10.1109/TCOMM.2019.2958916}
}

@article{mao2019rs_unicast_multicast,
  author  = {Yijie Mao and Bruno Clerckx and Victor O. K. Li},
  title   = {Rate-Splitting for Multi-Antenna Non-Orthogonal Unicast and Multicast Transmission: Spectral and Energy Efficiency Analysis},
  journal = {IEEE Transactions on Communications},
  volume  = {67},
  number  = {12},
  pages   = {8754--8770},
  year    = {2019},
  doi     = {10.1109/TCOMM.2019.2943168}
}
\end{document}